\documentclass[aps,prl,reprint]{revtex4-1}
\usepackage{blindtext}
\usepackage{graphicx}
\usepackage{subfigure}
\usepackage{dcolumn}
\usepackage{bm}
\usepackage[mathlines]{lineno}

\begin{document}
\title{Role of elasto-inertial turbulence in viscoelastic drag-reducing turbulence}
\author{Wenhua Zhang$^{1,2}$}
\author {Hongna Zhang$^{1}$}
\email{hongna@tju.edu.cn}
\author {Yuke Li$^{3}$}
\author{Bo Yu$^{4}$}
\author{Fengchen Li$^{1}$}
\email{lifch@tju.edu.cn}
\affiliation  {$^{1}$ Key Laboratory of Efficient Utilization of Low and Medium Grade Energy, MOE, School of Mechanical Engineering, Tianjin University, Tianjin 300350, China}
\affiliation {$^{2}$Sino-French Institute of Nuclear Engineering and Technology, Sun Yat-sen University, Zhuhai 519082, China}
\affiliation {$^{3}$Department of Physics of Complex System, Weizmann Institute of Science, Rehovot 7610001 Israel}
\address {$^{4}$School of Mechanical Engineering, Beijing Institute of Petrochemical Technology, Beijing 102617, China}

\date{\today}

\begin{abstract}
Two kinds of nonlinearities coexist in viscoelastic fluid flows, i.e., inertia and elasticity, which can engender different types of chaotic states including IT, DRT, ET and EIT. The state of MDR, the ultimate state of DRT induced by adding polymers to Newtonian IT, is recently regarded as EIT. This letter quantitatively demonstrates the role of IT and EIT in DRT via the contributions of RSS and the nonlinear part of ESS to flow drag. The nature of DRT is re-examined in a wide range of flow conditions covering a series of flow regimes from the onset of DR to MDR. We argue that EIT related dynamics appears in DRT long before settling to MDR regime and competitively coexists with IT in both spatial and temporal domain at moderate and high Re. More specifically, at low DR condition, EIT firstly emerges close to the channel walls. With the increase of elasticity, low-drag EIT gradually replaces high-drag IT from channel walls to center, resulting in a drastic decrease of flow drag comparing with IT. When EIT dynamics dominates the whole channel, MDR phenomenon is formed. Our findings provide evidences that DRT phenomenon is the result of IT and EIT interaction.
\end{abstract}

\maketitle

The unique rheology of viscoelastic fluid makes it show completely different flow behaviors from Newtonian fluid, such as drag-reducing turbulence (DRT) at moderate or high Reynolds number (Re) \cite{Toms1948} and elastic turbulence (ET) at extremely low Re \cite{Groisman2000}. These intriguing flow patterns inspire various new techniques for flow control in different industrial fields. For example, DRT by polymers, as a key technology, has been widely used in central heating/cooling systems, long-distance liquid transportation systems, and so on. ET has been excited to enhance heat and mass transfer at micro scale\cite{Yuan2020}. Due to the prospects of its wide application, viscoelastic fluid turbulence has become one of the hotspots in the field of fluid dynamics.

DRT has long been understood as a polymeric perturbation of Newtonian inertial turbulence (IT) \cite{White2008}. Recently the role of ET dynamics in DRT starts to be concerned \cite{Xi2019}. Elasto-inertial turbulence (EIT) is a newly discovered chaotic flow state observed in polymer solutions at high levels of elasticity \cite{Xi2019}. Since its discovery, a series of studies have been performed mostly focusing on its phenomenology, origin and implication on maximum drag reduction (MDR) \cite{Samanta2013}. In contrast to IT, EIT follows a different self-sustaining mechanism\cite{Dubief2013,Shekar2019}. Although a clear mechanistic understanding has not been reached yet so far, it is undoubted that elastic instability plays a dominant role. The distinctive features in EIT are that vortices are mostly in line with spanwise, originating from the localized sheet-like structures of high polymers extension \cite{Xi2019,Dubief2013,Zhu2019,Sid2018,Lopez2019}. Based on this point, EIT is regarded as two-dimensional (2D) turbulence and investigated through 2D simulations \cite{Zhu2019, Gillissen2019, Page2020}. Recent experiments and numerical simulations\cite{Choueiri2018,Xi2019,Lopez2019} also have demonstrated that MDR phenomenon, the ultimate state of DRT, is essentially a flow regime of EIT. The discovery of EIT and its relationship with MDR is a big breakthrough for viscoelastic fluid turbulence, especially for DRT. It not only harmonizes the seemingly contradictory results of both delayed turbulence \cite{Giles1967} and early turbulence \cite{Hansen1974} in DRT, but also provides direct proof for the existence of ET dynamics in DRT. The understanding of EIT is still at a preliminary stage for the short research history. Further excavation about its role in DRT (e.g., the flow dynamics near the wall in DRT), and how Newtonian turbulence develops into EIT (the connection between IT and EIT) are still necessities. 

The purpose of this letter is to draw a link between DRT and EIT by exploring the role of EIT in DRT. In wall turbulence, a large shear strain rate (SSR) forms in the near wall region where turbulent kinetic energy is pumped to the core region in IT, and provides a prerequisite for the excitation of EIT. On basis of experiments, Choueiri et al. speculated that there exists a “coexistence phase” of EIT and IT immediately preceding MDR\cite{Choueiri2018}. However, the underlying turbulent dynamics is hard to be determined using experimental means only. Siding with this speculation, we further conjectures that EIT dynamics could have been induced long before MDR in the near-wall region, which holds extremely large SSR, and even coexist with IT dynamics throughout DRT. With the increase of elasticity, EIT dynamics gradually expands to the whole channel, and finally dominates the turbulent flow. This letter provides numerical evidences for this conjecture. The key to verify this conjecture is to establish a quantitative characterization method for IT and EIT dynamics. Drag increase comparing with laminar flow is the most obvious feature of the flows accompanied with instability and turbulence. The stronger the related nonlinear effect is, the more the drag resistance increases. Based on this point, this letter quantifies the role of two dynamics in DRT by calculating the contribution of nonlinear Reynolds shear stress (RSS) and nonlinear part of elastic shear stress(ESS) to friction coefficient and then re-examines DRT in a wide range of flow conditions.

Direct numerical simulations (DNSs) are conducted for 3D plane Poiseuille flows of polymer solution under constant flow rate by Oldroyd-B model. The streamwise, wall-normal, and spanwise directions are denoted as $\it{x}$, $\it{y}$, and  $\it{z}$, with the corresponding velocity components are $\it{u}$, $\it{v}$ and $\it{w}$, respectively. Channel walls ($\it{y}$=0, 2$\it{\delta}$) are assumed to be non-slip and periodic boundary condition is applied in both the $\it {x}$ and $\it {z}$  directions with the periods,  $\it{L_x}$ and $\it{L_z}$. Taking the channel half height $\it{\delta}$, bulk mean velocity $\it{u_b}$,  $\it{\delta}/\it{u_b}$, $\it{\rho}/\it{u_b}$$^2$, as the reference length, velocity, time and pressure, the dimensionless governing equations are as follows: 

\begin{equation}
\frac{\partial {u^*_i}}{\partial {x^*_i}} = 0,
\end{equation}
\begin{equation}
\frac{\partial {u^*_i}}{\partial {t^*}} + {u^*_j}\frac{\partial {u^*_i}}{\partial {x^*_j}} = -\frac{\partial {p^*}}{\partial {x^*_i}} +\frac{2}{Re} \frac{\partial ^2 {u^*_i}}{\partial {x^*_i}^2} + \frac{\partial {\tau^*_{ij}}}{\partial {x^*_j}},
\end{equation}
\begin{equation}
\frac{\partial {c_{ij}}}{\partial {t}} + {u^*_k}\frac{\partial {c_{ij}}}{\partial {x^*_k}} ={c_{ik}}\frac{\partial {u^*_j}}{\partial {x^*_k}}+{c_{kj}}\frac{\partial {u^*_i}}{\partial {x^*_k}}-\frac{1}{\text{Wi}}(c_{ij}-\delta_{ij}),
\end{equation}
where $\it{\tau_{ij}}$ is the polymeric stress and is a function of the polymer conformation tensor $\it {c_{ij}}$, 
\begin{equation}
\tau^*_{ij} = \frac{2\beta(c_{ij}-\delta_{ij})}{\text{ReWi}};
\end{equation}
${\text{Re}}=2\it{\delta u_b/\nu}$ and ${\text{Wi}}=\it{\lambda u_b/ \delta}$ are Reynolds number and Weissenberg number; $\lambda$ is the relaxation time of polymer solution; $\beta=\eta/\nu$ is the viscosity contribution ratio ($\eta$ and $\nu$ are contribution of solute and solvent to zero shear viscosity of solution, respectively); $\delta_{ij}$ is the Kronecker function. The bounded second-order MINMOD scheme is employed to solve the convective term in Eq. (3) to guarantee numerical stability \cite{Yu2004}. Numerical details of DNS are described in supplementary material and verified in \cite{Zhang2020}. DNSs are performed in a wide range of flow conditions with $\it{\beta}$=1/9, Re=1000$\sim$20000, and Wi=0$\sim$60 (Wi=0 denotes Newtonian fluid flow) covering a series of flow regimes including laminar flow, IT, EIT and DRT from the onset of DR to MDR. As the near-wall low-speed streaks and polymer extension sheet-like structures elongates with the increase of Wi, the channel length and width $\it{L_x}$ and $\it{L_z}$ are enlarged for higher Wi to capture those structures during the simulations. The spatial resolution of numerical simulation is $256\times128\times128$, and the dimensionless time step is $\Delta t^*$=0.002.

As nonlinear inertia is responsible for IT and nonlinear elasticity for EIT, this letter compares the role of IT and EIT in viscoelastic DRT via the contributions of RSS and the nonlinear part of ESS to flow drag. Taking the nature of viscoelastic DRT as IT, conventional studies explain the mechanism of DRT focusing on the reduction of Reynolds stress induced by nonlinear elasticity. In contrast with RSS, the formation mechanism of ESS and its role in the flow drag are paid much less attention. Based on Eqs. (3) and (4), mean ESS in viscoelastic turbulent flows can be decomposed into (see the derivation details in supplementary material),

\begin{eqnarray}
\left\langle {\tau _E^* } \right\rangle  &=& \underbrace {\frac{{2\beta }}{{{\text{Re}}{_b}}}\frac{{\partial \left\langle {{u^* }} \right\rangle }}{{\partial {y^* }}}}_{{A}}  + \underbrace {{\text{Wi}}\left( {\left\langle {\tau'^{*}_{k2}\frac{\partial {u'^{*}}}{\partial {x^*_k}}} \right\rangle  + \left\langle {\tau'^{*}_{k1}\frac{\partial {v'^{*}}}{\partial {x^*_k}}} \right\rangle } \right)}_{{B}} \nonumber \\
& & + \underbrace {{\text{Wi}}\left\langle {\tau _{22}^ * } \right\rangle \frac{{\partial \left\langle {{u^ * }} \right\rangle }}{{\partial {y^ * }}}}_{{C}}-\underbrace {{\text{Wi}}\left\langle {u'^{*}_k}\frac{\partial {\tau'^{*}_E}}{\partial {x^*_k}} \right\rangle }_{{D}}
\end{eqnarray}
where $\left\langle {\cdot} \right\rangle$ is the symbol of ensemble average; $\it{A}$ is a linear term proportional to SSR; $\it{B}$, $\it {C}$ and $\it{D}$ are nonlinear terms which are zero in laminar flow. Obviously, mean ESS in viscoelastic turbulent flows consists of a linear mechanism of the increase of apparent viscosity ($\it{A}$), and a nonlinear mechanism closely related to the interaction between polymers and turbulence. To be concise, we decompose ESS into two parts, $\left\langle {\tau _E^ * } \right\rangle {\rm{ = }}\left\langle {\tau _{E1}^ * } \right\rangle {\rm{ + }}\left\langle {\tau _{E2}^ * } \right\rangle$ with $\left\langle {\tau^*_{E1}}\right\rangle=A$, and $\left\langle {\tau^*_{E2}}\right\rangle=B+C-D$. 

To demonstrate the roles of RSS and the nonlinear part of ESS in DRT, Renard-Deck (RD) identity is introduced \cite{Zhang2020}:
\begin{eqnarray}
{C_f} &=& \underbrace {\int_0^2 {\left\langle {\tau _V^ * } \right\rangle  \cdot Sd{y^ * }} }_{{C_{f,{\rm{ }}V}}} + \underbrace {\int_0^2 {\left\langle {\tau _R^ * } \right\rangle  \cdot Sd{y^ * }} }_{{C_{f,{\rm{ }}R}}} \nonumber\\
& &+ \underbrace {\int_0^2 {\left\langle {\tau _{E1}^ * } \right\rangle  \cdot Sd{y^ * }} }_{{C_{f,{\rm{ }}E1}}} + \underbrace {\int_0^2 {\left\langle {\tau _{E2}^ * } \right\rangle  \cdot Sd{y^ * }} }_{{C_{f,{\rm{ }}E2}}}
\end{eqnarray}
where $S = {{\partial \left\langle {{u^ * }} \right\rangle } \mathord{\left/{\vphantom {{\partial \left\langle {{u^ * }} \right\rangle } {\partial {y^ * }}}} \right.
 \kern-\nulldelimiterspace} {\partial {y^ * }}}$ is the mean SSR; $\left\langle {\tau^*_V} \right\rangle =2S/{\text{Re}}$ is the mean viscous shear stress (VSS); $\left\langle {\tau^*_R} \right\rangle =-\left\langle {u'^{*}v'^{*}} \right\rangle $ is the mean RSS; $C_{f,V}$, $C_{f,R}$, $C_{f,E1}$ and $C_{f,E2}$ are the contributions of viscous, Reynolds, linear elastic, and nonlinear elastic shear stresses to the flow drag, respectively. 

\begin{figure}
\centering
\subfigure
{
  \begin{minipage}[b]{\linewidth}
        \centering
        \includegraphics[width=0.6\textwidth]{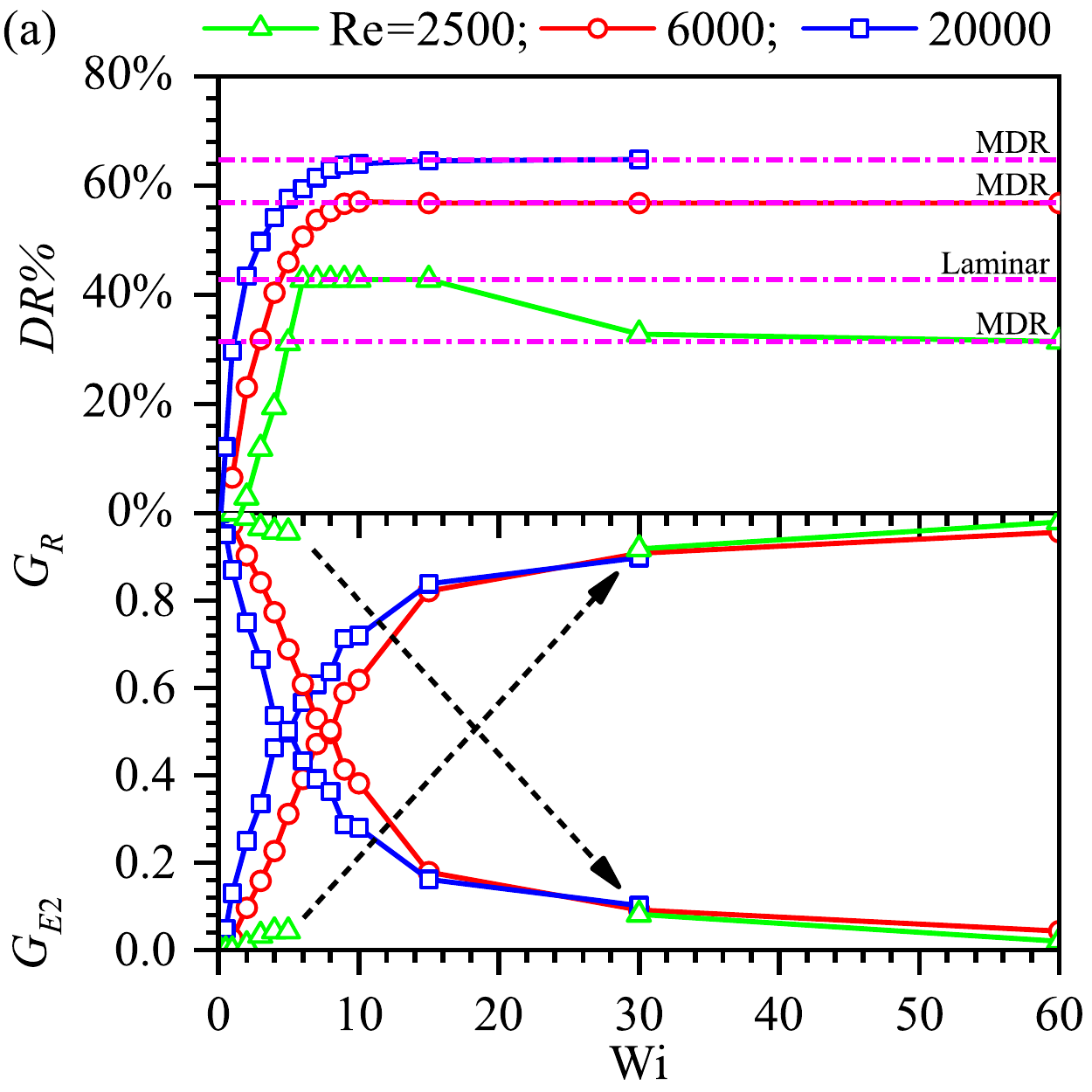}
    \end{minipage}
}
\subfigure
{
  \begin{minipage}[b]{\linewidth}
        \centering
        \includegraphics[width=0.8\textwidth]{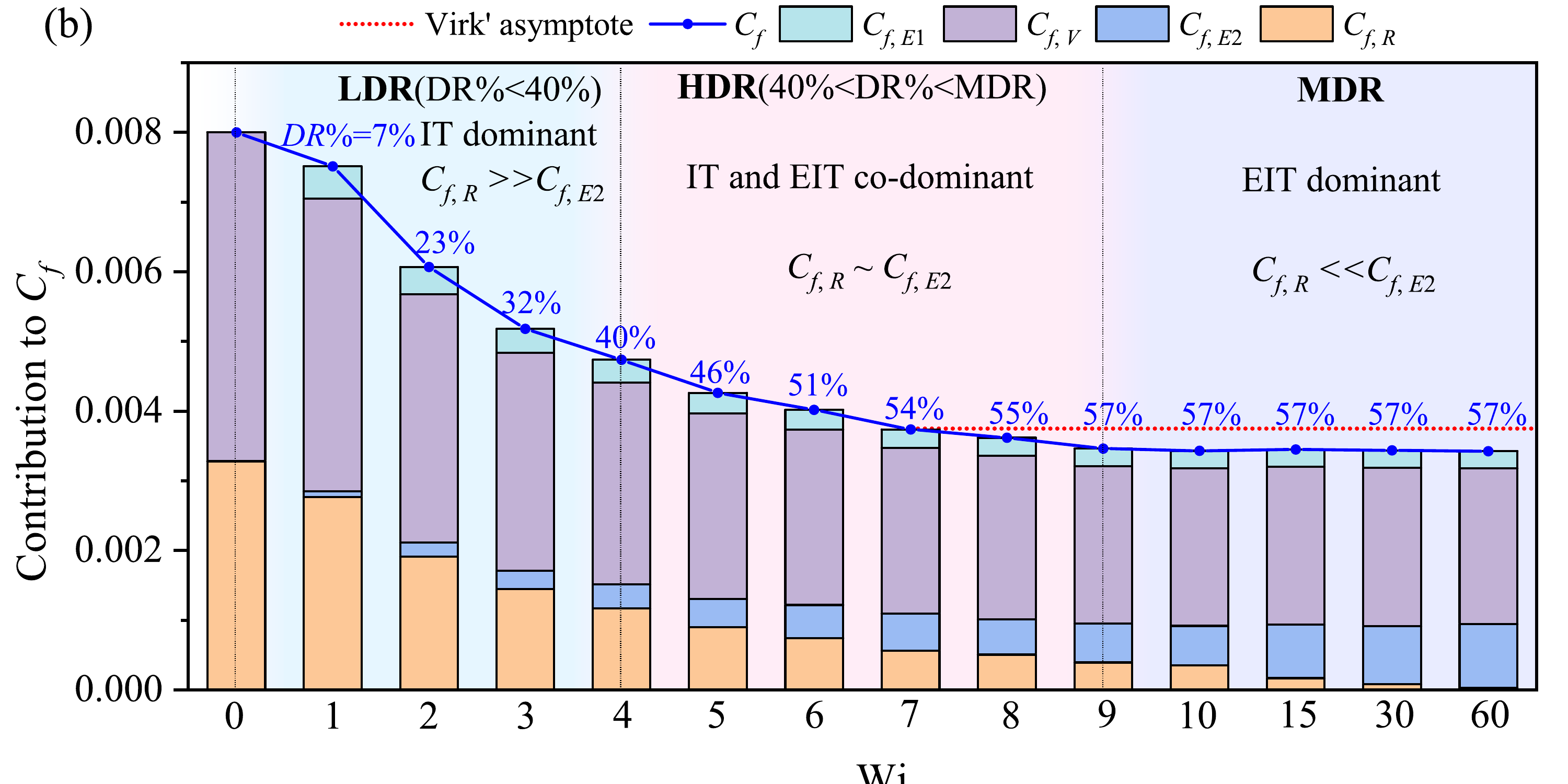}
    \end{minipage}
}
\caption{\label{fig1} Distributions of (a) $G_{E2}$ and $G_R$ with DR ratio at different Re and Wi; (b) $C_f$ and its contributions with DR ratio at Re=6000. $G_{E2}$ and $G_R$ represent the ratio of nonlinear elastic effect and inertial effect to the total, which are calculated by $G_{E2}=C_{f,E2}/( C_{f,E2}+ C_{f,R})$ and $G_R=C_{f,R}/( C_{f,E2}+ C_{f,R})$, respectively. $DR\%=(C_{f}^N- C_f^V)/C_{f}^N\times100\%$. '$N$'denotes Newtonian fluid and '$V$'denotes viscoelastic fluid. Virk's asymptote is calculated by $1/C_f^{0.5}=19\rm{lg}({\text{Re}}_b{\it{C_f}}^{0.5})-32.4$.}
\end{figure}

We firstly demonstrate the applicability of using $C_{f,E2}$ and $C_{f,R}$ to make a quantitative comparison of the role of EIT dynamics and IT dynamics, respectively, in viscoelastic turbulence. Contributions of different stresses to drag coefficients $C_f$ and drag modification by elasticity are illustrated and compared for Newtonian and viscoelastic fluid flows in Fig.\ref{fig1}a. At low Re (e.g., 1000, Newtonian fluids: laminar), viscoelastic fluid flow becomes unstable and EIT dominates when $Wi > Wi_c$. For a small Re (e.g., 2500, Newtonian fluids: IT), there are two stages of DR. With increase of Wi, viscoelastic fluid flow exceeds MDR asymptote and enters a laminar regime (Stage I: Relaminarization, at Re = 2500 and Wi in between 6 and 15 as shown in Fig. \ref{fig1}a) and settles to EIT with further increase of Wi (Stage II). The occurrence of laminar regime draws a clear distinction between IT and EIT, in consistent with Choueiri’s experimental observations \cite{Choueiri2018}. At moderate Re (e.g., 6000 or 20000, Newtonian fluids: IT), common to earlier computations, $C_f$ decreases gradually to just below Virk’s asymptote with the increase of Wi and saturates to form MDR phenomenon with DR$\%$ of 57$\%$ for Re = 6000 and of 64$\%$ for Re = 20000. Striking in the above different cases is the manifestation of $C_{f,E2}$, which is negligible in IT, but prominent in EIT. In cases of IT and before the onset of DR, as the nature of turbulence is inertial, $C_{f,R}$ is the only mechanism responsible for the additional flow drag beyond laminar flow with $C_{f,E2}=0$. In MDR stage, nonlinear elastic stress replaces the nonlinear inertial one and is mainly responsible for the additional flow drag comparing with the laminar flow, i.e., $C_{f,R} << C_{f,E2}$. With further increase of Wi, IT is eventually marginalized and EIT with  $C_{f,R} = 0$ takes over inconsistent with the mainstream view so far that IT is replaced by EIT at MDR stage \cite{Samanta2013,Choueiri2018}. Therefore, it is indeed reliable to use  $C_{f,E2}$ to characterize EIT dynamics, just as $C_{f,R}$ to represent IT dynamics. On account of this quantitative characterization, this letter concentrates on a representatively moderate Re (Re = 6000, Fig. 1b), which is often met in the related DRT investigations. Similar conclusions are also achieved for the condition of Re= 20000 (not shown for brevity).

At a moderate and high Re, viscoelastic DRT is limited by the so-called MDR asymptote. In between IT and EIT, it is impressive to see that, in contrast with $C_{f,R}$, $C_{f,E2}$ shows a drastic increase with Wi (Fig.\ref{fig1}a and Fig.\ref{fig1}b). It implies that $\tau_{E2}$ starts to contribute to $C_f$ long before entering MDR. At a high-extent drag-reduction (HDR) condition (Wi=8),$C_{f,E2}$ catches up with  $C_{f,R}$, that is, EIT generally begins to play an equally important role as IT. Hereafter, EIT and IT exchange their roles in the turbulent flow dynamics. Therefore, we argue that EIT dynamics and IT dynamics coexist with each other even throughout DRT. At low-extent drag-reduction (LDR) stage, turbulence fluctuations mainly serve the maintenance of IT, while at MDR stage mainly participates in the maintenance of EIT. But the coexistence and competitive relationship between EIT and IT should be more complicated, considering different characteristics and dynamics of LDR, HDR and MDR.

\begin{figure}[htbp]
\centering
\subfigure
{
  \begin{minipage}[b]{\linewidth}
	   \centering
        \includegraphics[width=0.8\textwidth]{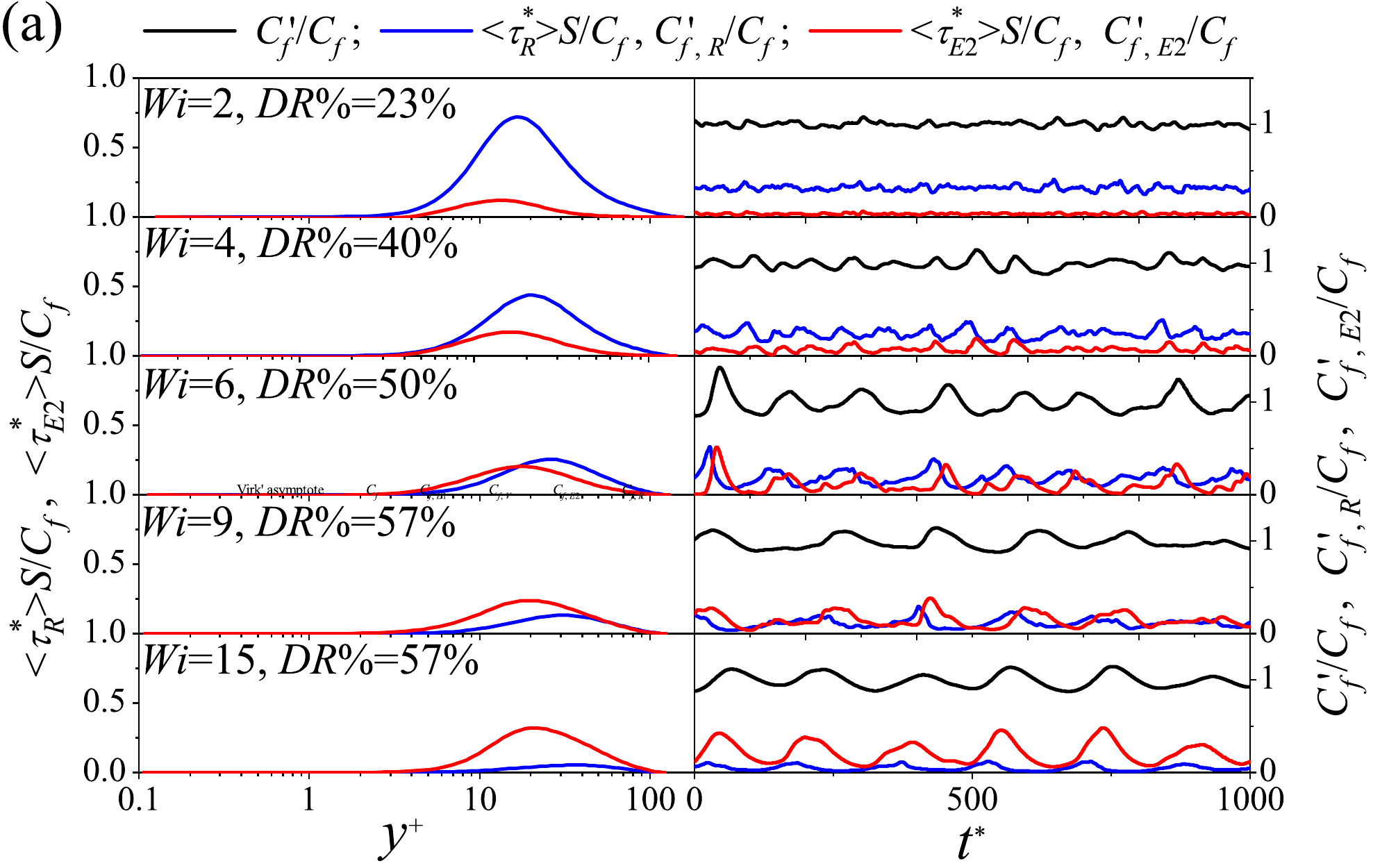}
    \end{minipage}
}
\subfigure
{
  \begin{minipage}[b]{\linewidth}
	   \centering
        \includegraphics[width=0.8\textwidth]{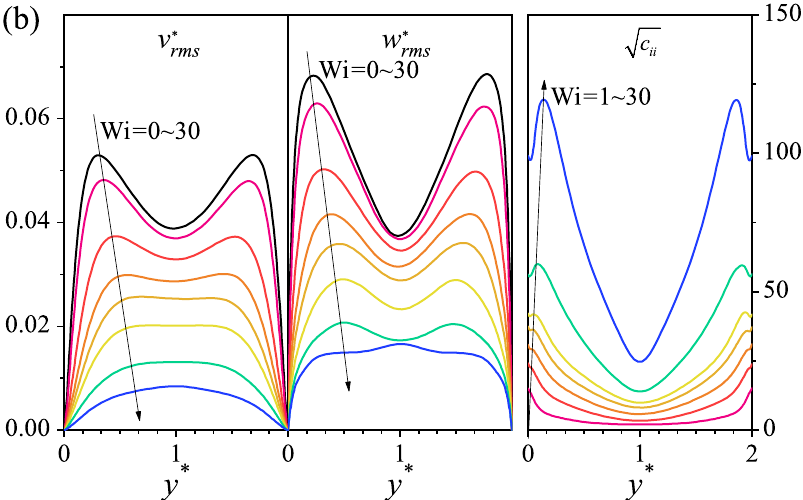}
    \end{minipage}
}
\caption{\label{fig2} (a) Distributions of integrand of $ C_{f,R}/C_f$ and $C_{f,E2}/C_f$ as a function of $y^+$ (left) and evolutions of unified instantaneous friction coefficients (right); (b) distributions of $v^*_{rms}$, $w^*_{rms}$, $\sqrt{c_{ii}}$ with respect to $y^*$. Here, the superscript prime denotes the instantaneous variable after spatial averaging. }
\end{figure}

\begin{figure}
\includegraphics[width=0.4\textwidth]{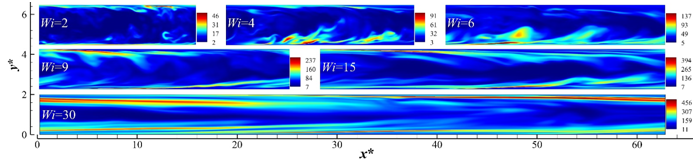}
\caption{\label{fig3} Instantaneous distributions of $\sqrt{c_{ii}}$ in the $x-y$ plane.}
\end{figure}

Figure \ref{fig2}(a) plots the spatial (left) and temporal (right) statistical characteristics of $C_{f,E2}$ and $C_{f,R}$. It can be seen from the spatial distributions that, with increase of Wi, $C_{f,E2}$ is firstly lifted in the near-wall region, gradually becoming comparable to $C_{f,R}$, and then to the core region, finally far exceeding $C_{f,R}$ in the whole channel. The underlying physical process could be as follows. EIT firstly occurs close to the wall, and favors buffer layer with high SSR just like IT; with the increase of elasticity, EIT dynamics is enhanced and IT dynamics is forced to retreat towards the channel center accompanied by that the peak of $C_{f,R}/C_f$ gradually decreases and moves up for its position. In response, peaks of profiles of $v^*_{rms}$ and $w^*_{rms}$ gradually shift towards the channel center until their concavity disappears (Fig.\ref{fig2}b). Coherent structures related with IT are gradually lifted to the channel center when EIT dynamics gets involved from near-wall region and finally vanishes therein when EIT dynamics occupies the whole channel \cite{Dubief2013} (See the distributions of streamwise vortices extracted in $y-z$ plane in Figs. S4 in supplementary material). More specifically, at a LDR condition (e.g., Wi=2), EIT just rises and is not enough to shake the dynamic dominance of IT. For this reason, the weakened flow characteristics of IT with “thickened buffer layer” can usually be observed at LDR condition. At a HDR condition (e.g., Wi=6), an interesting situation of zoning happens, that is, IT dynamics can still occupy the turbulent core region while the pivotal near-wall dynamics is usurped by EIT. This is the essential reason that HDR is distinguished with LDR \cite{Xi2019}. At a MDR condition (e.g., Wi=15), EIT dynamics fully defeats IT dynamics in the whole channel region. As one of the most distinctive features of EIT, sheet-like structures of high polymer extension \cite{Xi2019,Dubief2013,Zhu2019,Sid2018,Lopez2019} can be clearly observed at the LDR condition (e.g., Wi = 2) and gradually expand towards channel center with the increase of Wi (Fig.\ref{fig3}). In statistics, the hump-like profile of $\sqrt {c_{ii}}$ starts to be formed in LDR and gets enhanced with Wi (Fig.\ref{fig2}b). As for the temporal characteristics, under a LDR condition (e.g., Wi=2), the role of EIT is always weaker than that of IT, i.e.,$C'^{}_{f,E2}<<C'^{}_{f,R}$, and $C'^{}_{f}$ retains the random characteristics like IT with a smaller characteristic time scale. Contrarily, under a condition of MDR (e.g., Wi=15), EIT dynamics dominates the temporal evolution, i.e., $C'^{}_{f,E2}>>C'^{}_{f,R}$, and $C'^{}_{f}$ becomes almost quasi-periodic with a much larger characteristic time scale. In between them, moments of EIT being dominant appear more and more frequently, and $C'^{}_f$ becomes more and more regular as a result. Also interesting is that there exists a remarkable phase shift between  $C'^{}_{f,E2}$ and $C'^{}_{f,R}$ (about 1/4 quasi-cycle obtained through correlation analysis) since the flow enters HDR regime (Wi = 4), implying interaction somehow between IT and EIT. Based on these findings, we argue that EIT dynamics gradually replaces IT dynamics from the near-wall region to the turbulent core region throughout DRT.

\begin{figure}[htbp]
\centering
\subfigure
{
  \begin{minipage}[b]{.45\linewidth}
        \centering
        \includegraphics[width=\textwidth]{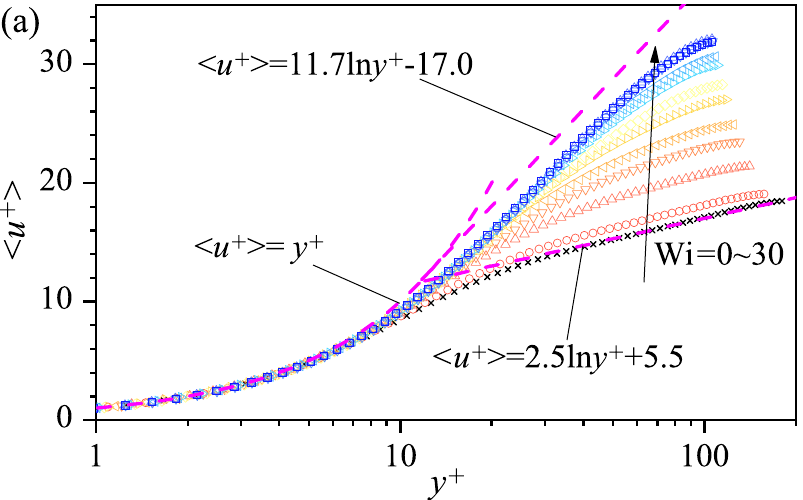}
    \end{minipage}
}
\subfigure
{
  \begin{minipage}[b]{.45\linewidth}
        \centering
        \includegraphics[width=\textwidth]{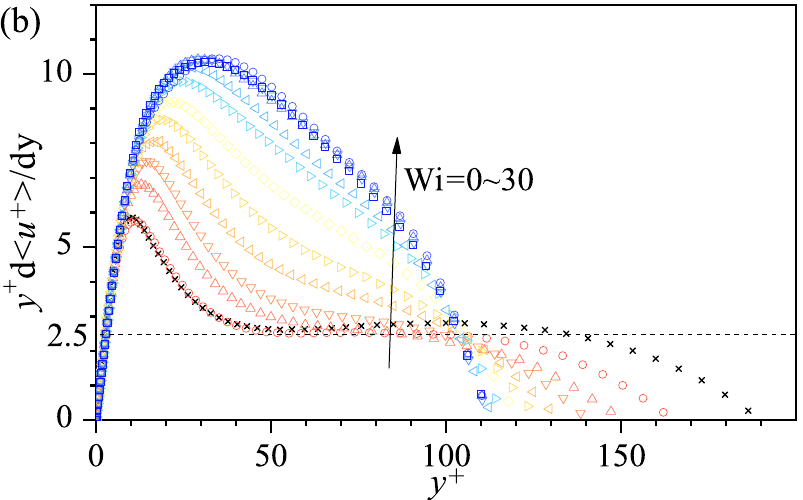}
    \end{minipage}
}
\subfigure
{
  \begin{minipage}[b]{.45\linewidth}
        \centering
        \includegraphics[width=\textwidth]{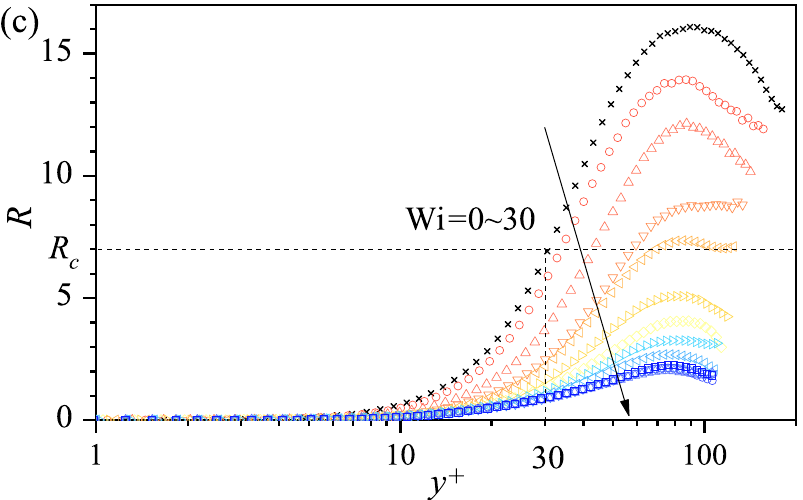}
    \end{minipage}
}
\subfigure
{
  \begin{minipage}[b]{.45\linewidth}
        \centering
        \includegraphics[width=\textwidth]{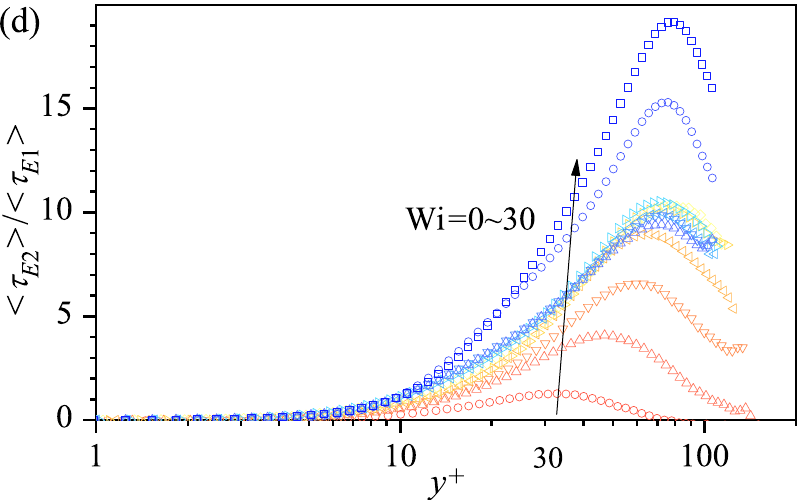}
    \end{minipage}
}
\caption{\label{fig4} Profiles of (a)$\left\langle {u^+}\right\rangle$ ; (b) the slope of $\left\langle {u^+}\right\rangle$; (c) ratio of the contributions from nonlinear terms to linear terms $R=(\left\langle {\tau _R^*} \right\rangle + \left\langle {\tau _{E2}^*} \right\rangle)/(\left\langle {\tau _V^*} \right\rangle + \left\langle {\tau _{E1}^*} \right\rangle)$;(d) $ \left\langle {\tau _{E2}^*} \right\rangle/\left\langle {\tau _{E1}^*} \right\rangle$ at different Wi.}
\end{figure}

At last, we use the above findings to re-evaluate the mean velocity $\left\langle {{u^ + }} \right\rangle$ profile (as shown in Fig.\ref{fig4}a). The shape of $\left\langle {u^+}\right\rangle$ profile is a manifestation of the underlying flow dynamics. The existence of log-law layer with slope of 2.5 (at $y^{+}>30$) is universal and typical features for IT \cite{Pope2000}. In log-law layer, nonlinear inertial effect ($C_{f,R}$) is far larger than linear viscous effect ($C_{f,V}$), and there exists a critical value ($R_c \approx 7$ at Re=6000) for ratio of nonlinear effect to linear effect ($R =C_{f,R}/C_{f,V}$ for IT) at  $y^{+}\approx 30$ (see Fig.\ref{fig4}c), which is often considered as the border between the log-law layer and buffer layer. Although being questioned recently \cite{White2018}, it is still undeniable that Virk's asymptote closely describes the profile of $\left\langle {{u^+}} \right\rangle$  in MDR \cite{Xi2019} where EIT dynamics dominates the whole channel, and can be used as an indicator of EIT dynamics. At LDR stage (e.g., Wi =2), EIT dynamics starts to occur in the near-wall region, but weaker than IT dynamics among the whole channel (Fig.\ref{fig2}a). The profile of $\left\langle {{u^+}} \right\rangle$  is lifted up towards the law of EIT in the near-wall region, but still satisfying the log law of IT in the core region (Fig.\ref{fig4}b). The ratio $R$ of nonlinear effects ($C_{f,R}$ and $C_{f,E2}$) to linear effects ($C_{f,V}$ and $C_{f,E1}$) is delayed to reach $R_c$ (Fig. \ref{fig4}c), resulting in a shortened quasi-flat log-law layer compared with that in IT (Fig. 4a). At HDR stage (e.g., Wi = 6), EIT dynamics exceeds that of IT in the near-wall region at  $y^{+}\approx 20$  or further and is comparable in the core region (Fig.\ref{fig2}a). As a result, the profile of $\left\langle {{u^+}} \right\rangle$  is collapsed onto the law of EIT in the near-wall region, and apparently deviates from log law of IT (or in between the law of IT and EIT) in the core region. Unlike that in LDR, $R$ can never reach $R_c$ since entering HDR (see Fig.\ref{fig4}c). The hypothesis to derive the log law in IT fails and the quasi-flat log-law layer is completely eradicated (see Fig. 4a). At MDR stage, $R$ converges to an asymptote far below $R_c$ in the whole channel. Notable in Fig.\ref{fig4}c is that the maximum value of $R$ asymptote is about 2.5, comparable to that in the buffer layer for IT. Under this condition, the linear viscous effects cannot be ignored for the whole channel area. Rather than a thicker IT buffer layer or weak IT dynamics, the flow is, however, finally replaced by a low-drag mode of nonlinear dynamics (EIT) with significantly amplified nonlinear elastic effect (see Fig.\ref{fig4}d). 

In summary, we have quantitatively characterized the roles of IT and EIT dynamics in viscoelastic turbulent channel flow by means of their contributions to the flow drag resistance. In this framework, we have pictured the flow characteristics and mechanisms at different DR stages, and drawn a link between viscoelastic DRT and EIT. Under small-Re condition, in consistent with Choueiri’s experiments \cite{Choueiri2018}, with the increase of Wi, the viscoelastic flow enters laminar regime before EIT is excited exceeding MDR asymptote. Starting from laminar regime, viscoelastic flow enters EIT with further increase of Wi. Under moderate and high Re condition (e.g., 6000 or 20000), our results demonstrated that EIT dynamics starts to get involved long before entering MDR, even throughout DRT; with the increase of Wi, EIT dynamics is enhanced and the turbulent flow is gradually dominated by EIT dynamics from near-wall region to turbulent core region; when EIT dominates the whole channel area, MDR phenomenon occurs. Finally, we argue that polymers have dual effects in DRT: modulating IT dynamics and introducing additional EIT dynamics. These two effects lead to two different DR patterns (Fig. \ref{fig1}a): flow re-laminarization for small Re cases (the nature of turbulence is still IT) and the replacement of high-drag mode turbulence by low-drag mode turbulence (the nature of turbulence gradually changes from IT to EIT) for moderate or high Re.

This research was funded by the National Natural Science Foundation of China (NSFC 51976238, 51776057, 52006249).
\bibliographystyle{apsrev4-1} 

\end{document}